\documentclass[preprint,aps]{revtex4}

 \usepackage{amssymb}
 \usepackage{graphicx}
 \usepackage{latexsym}
 \usepackage{amsmath}
 \usepackage{amsfonts}
 \usepackage{verbatim}
 \usepackage{amsthm}
 \usepackage{epsfig}
 \usepackage{graphics}
 
\newcommand {\tauin}{\tau_{\rm{in}}}
\newcommand {\tauout}{\tau_{\rm{out}}}
\newcommand {\tauopen}{\tau_{\rm{open}}}
\newcommand {\tauclose}{\tau_{\rm{close}}}
\newcommand {\vcrit}{v_{\rm{crit}}}

\newcommand {\Dstar}{D^{*}}
\newcommand {\Astar}{A^{*}}
\newcommand {\Gstar}{G^{*}}

\begin{document}

\title{Rate-dependent propagation of cardiac action
       potentials in a one-dimensional fiber}
\author{
John W. Cain\footnote{Electronic mail: jcain@math.duke.edu},$^{1}$
\ Elena G. Tolkacheva,$^{2}$\ David G. Schaeffer,$^{1,4}$
and Daniel J. Gauthier$^{2,3,4}$}

\address{
$^{1}$Department of Mathematics, \medskip $^{2}$Department of Physics, \\
$^{3}$Department of Biomedical Engineering, and $^{4}$Center for Nonlinear and
Complex Systems,\\
Duke University, Durham, North Carolina 27708\bigskip , USA}
\date{\today}

\begin{abstract}
Action potential duration (APD) restitution, which relates APD to
the preceding diastolic interval (DI), is a useful tool for predicting
the onset of abnormal cardiac rhythms.  However, it is known that 
different pacing protocols lead to different APD restitution curves (RCs).
This phenomenon, known as APD rate-dependence, is a consequence of memory
in the tissue.  In addition to APD restitution, conduction velocity 
restitution also plays 
an important role in the spatiotemporal dynamics of cardiac tissue.
We present new results concerning rate-dependent restitution in the velocity 
of propagating action potentials in a one-dimensional fiber.  
Our numerical simulations show that, independent of the amount of memory
in the tissue, waveback velocity exhibits pronounced rate-dependence and 
the wavefront velocity does not.  Moreover, the discrepancy between waveback 
velocity RCs is most significant for small DI.  We provide an analytical 
explanation of these results, using a system of coupled maps to relate the 
wavefront and waveback velocities.  Our calculations show that waveback 
velocity rate-dependence is due to APD restitution, not memory.

\end{abstract}

\maketitle
 
\section{INTRODUCTION}
 
When a cardiac cell is depolarized by an electrical stimulus, it 
exhibits a prolonged elevation of transmembrane potential known as an 
action potential.  We define the action potential duration 
(APD) as the time required for the cell to achieve 80\% repolarization 
following a depolarizing stimulus.  The refractory period 
between the end of an action potential the application of a subsequent 
stimulus is called the diastolic interval (DI).  It is known that APD 
restitution (the dependence of the APD on preceding DI) is of fundamental 
importance in paced cardiac dynamics. In particular, studies 
\cite{ND, GUEVARA} show that the slope of the APD restitution curve (RC) 
is linked to the onset of alternans, an abnormal cardiac rhythm 
characterized by long-short variation of APD, which may lead to 
ventricular fibrillation and sudden cardiac death
\cite{KARMA94, ROSENBAUM, WATANABE-OG}.
 
Experimental \cite{BOYETT, ELHARRAR, FRANZ} and analytical \cite {TSGK} 
investigations have shown that different pacing protocols
lead to different APD RCs, a phenomenon known as APD rate-dependence. 
Several studies \cite{GULRAJANI, CHIALVO, OTANI} indicate that the origin 
of APD rate-dependence is the 
presence of memory in cardiac tissue. That is, APD depends not only upon 
the preceding DI but also on the previous history of paced cardiac tissue. 
Memory appears to be a generic feature of cardiac muscle since it has been 
reported in humans \cite{FRANZ} and various animals 
\cite{ELHARRAR, HBG, GETTES, GIBBS}.
 
Testing for rate-dependence involves the use of multiple pacing
protocols and comparison of the resulting RCs.  Two of the most commonly used 
pacing schemes are the
dynamic and S1-S2 pacing protocols. Under the dynamic (steady-state)
protocol, pacing is performed at a constant basic cycle length $B$ 
until steady-state is reached (no beat-to-beat variation in APD or DI).
After recording the steady-state
DI-APD pair, $B$ is changed by an amount $\Delta$ and the
process is repeated. The dynamic RC is constructed by
plotting all steady-state DI-APD pairs obtained from the dynamic
pacing protocol over a range of $B$ values. The S1-S2 (standard) 
protocol also begins with pacing at a fixed basic cycle length $B$ 
(S1 interval) until steady-state is reached. Then, an S2 stimulus is applied 
at an interval $B_{1}$ after the final S1 stimulus. Setting 
$\delta = B_{1}-B$, the S1-S2 RC is obtained by plotting the APD following 
the S2 stimulus versus the preceding DI for different values of $\delta$. 
Note that there is only one dynamic RC, whereas each different S1 pacing 
interval can yield a distinct S1-S2 RC. For the purposes of this paper, we 
obtain
only {\em local} S1-S2 RCs ($|\delta|$ small relative to S1) for 
different values of the S1 interval. In particular, following \cite{KALB} 
we apply one short $(B_{1} = B-\delta)$ and one long $(B_{1} = B+\delta)$ 
perturbation at each different value of $B$. 

\begin{figure}[tbp]
 
 
  \includegraphics[width=8cm]{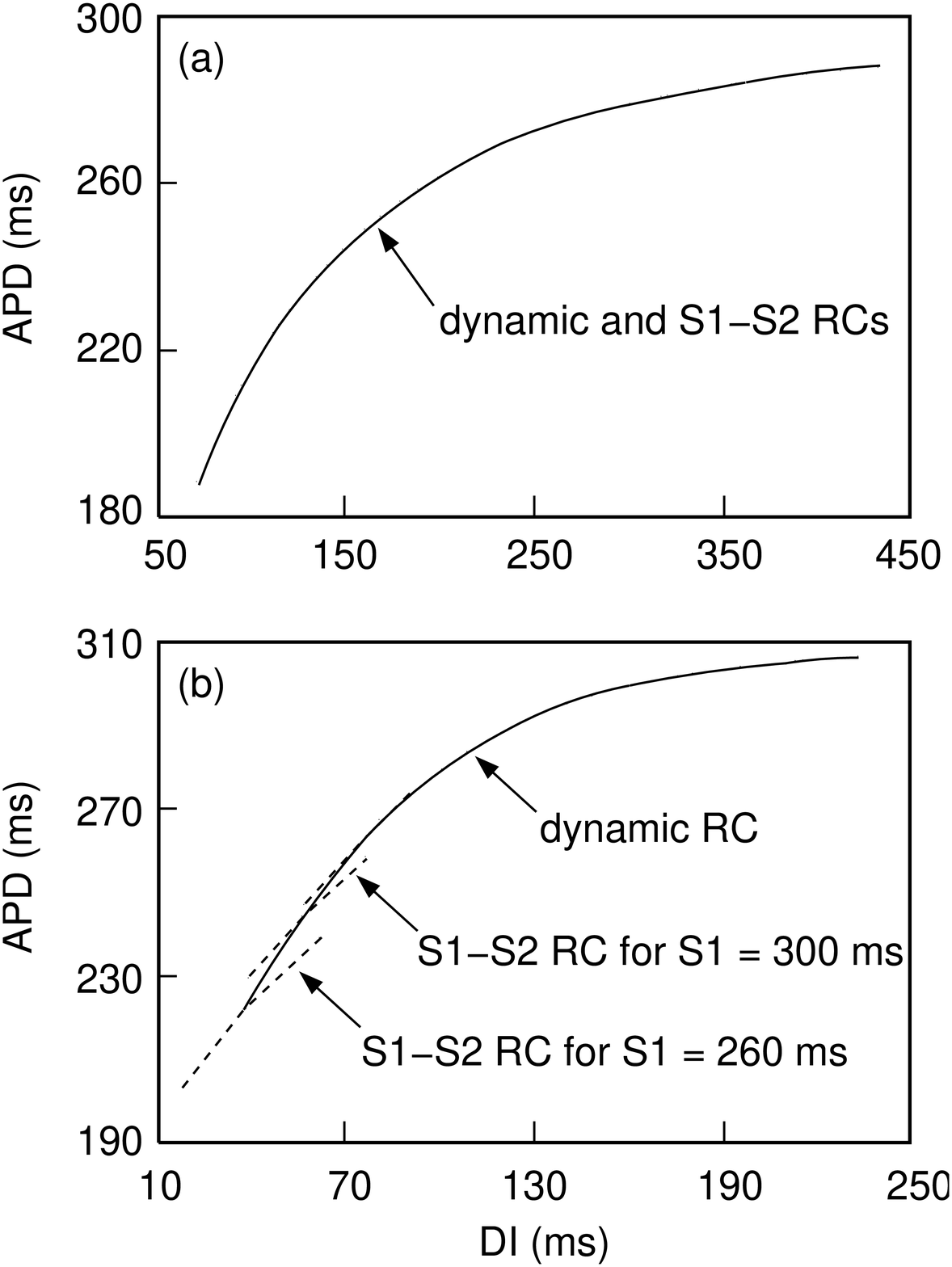}
 
  \caption{Typical APD restitution curves obtained using dynamic and 
  S1-S2 pacing 
  protocols ($\Delta = 40$ ms, $\delta = \pm 20$ ms) for two different ionic 
  membrane models. 
  (a) A two-current model. 
  (b) A three-current model. 
  The dynamic RC (solid) and local S1-S2 RCs (dashed) are shown for
  different values of S1.
  }
 
  \label{COMPARE-APD}
 
\end{figure}

The connection between rate-dependent APD restitution and memory is
illustrated in Fig. \ref{COMPARE-APD}, which
shows APD RCs obtained from numerical 
simulations using two different ionic models (see Section II) of the cell 
membrane.  Figure \ref{COMPARE-APD}a is generated using a two-current 
model \cite{KARMA93, MS} with no memory:  the dynamic and S1-S2 RCs are 
indistinguishable.  Figure \ref{COMPARE-APD} is generated using a 
three-current ionic model \cite{FK, TSGM} with some memory.  One can see from 
Fig. \ref{COMPARE-APD}b that segments of S1-S2 RCs (dashed curves) do not 
coincide with the dynamic RC (solid curve) for small DI values.  The 
splitting between the dynamic and S1-S2 RCs is the manifestation of APD 
rate-dependence.
 
In the case of a network of electrically-coupled cardiac cells, 
conduction-velocity restitution also plays an important role in dynamics of 
the spatially extended system \cite{BANVILLE, CKG}. Indeed, when a sequence 
of propagating 
pulses is produced, both APD and the propagation speed of a pulse are
influenced by the preceding pulse. Conduction-velocity restitution is 
analogous to APD restitution in that it relates the speed of an action
potential at a given site to the DI at that site.  Several authors
\cite{CKG, COURTEMANCHE, FENTON-CHE} have noted that abrupt changes in 
the pacing rate lead to discrepancies between the wavefront and waveback
velocities.  For this reason, we will always distinguish between
wavefront (or activation front) velocity and waveback (or recovery front)
velocity of propagating action potentials.

In this paper, we investigate rate-dependence of wavefront and 
waveback velocities of propagating action potentials in a one-dimensional 
fiber of cardiac cells.  Using numerical simulations of different ionic
membrane models (with and without memory) we demonstrate that
the waveback velocity exhibits pronounced rate-dependence and the wavefront 
velocity does not (Section II).  We derive an analytical relationship between 
wavefront and waveback velocities by modeling cardiac dynamics using a 
system of coupled maps with an arbitrary amount of memory.  We show that
APD restitution, not memory, leads to waveback velocity
rate-dependence (Section III).  We provide conclusions and discussion in 
Section IV.  An Appendix on the two-current ionic model is included for
reference.
 
 
\section{Rate-dependent velocity:  numerical results}
 
Typically, the cardiac action potential is modeled by considering ionic
currents that flow across the cell membrane via ion channels.
The rate-of-change of the transmembrane voltage is obtained by summing all 
ionic currents and dividing by the membrane capacitance. The ion channels 
act as gates that regulate the permeabilities of ions, most notably sodium, 
potassium and calcium.  Hence, ionic models are presented as systems 
of ordinary differential equations that govern transmembrane voltage and 
gate variables.

In the case of a one-dimensional fiber, electrical coupling can be modeled
by the inclusion of a diffusion term.  The result is a 
reaction-diffusion partial differential equation known as the cable 
equation:
\begin{equation} 
  \frac{\partial v}{\partial t} = \kappa \frac{\partial^{2} v}{\partial x^{2}}
  - \frac{I_{\mathrm{total}}}{C_{m}} \hspace{1.0 true cm} 0 \leq x \leq L,
  \label{CABLE}
\end{equation}
where $v$ denotes transmembrane voltage, $x$ measures distance from the
stimulus site, $C_{m}$ is membrane capacitance, $\kappa$ is a diffusion
coefficient, and $I_{\mathrm{total}}$ is the sum of all ionic currents. 
The number of currents varies depending upon the complexity of the ionic
model.  The diffusion coefficient incorporates membrane capacitance, cell
surface-to-volume ratio, and longitudinal resistivity of cardiac muscle
tissue. In all of our numerical simulations 
$C_{m}=\mathrm{1 \: \mu F \: cm^{-2}}$
and $\kappa=\mathrm{0.001 \: cm^{2} \: ms^{-1}}$. Neumann boundary conditions 
$v_{x} (0, t) = v_{x}(L, t) = 0$ are imposed at both ends of the cable.

To investigate rate-dependent propagation, we perform numerical simulations of
Eq. \eqref{CABLE}.  We apply both dynamic and S1-S2 pacing protocols at one
end of a cable, and measure the wavefront and waveback velocities of each
propagating pulse.  By analogy with APD rate-dependence, velocity 
rate-dependence means that different pacing protocols lead to different 
velocity RCs.

Since memory is responsible for APD rate-dependence, it is natural to 
hypothesize that memory also leads to wavefront and waveback velocity 
rate-dependence.  Consequently, we use two different ionic models in our
numerical simulations: a two-current ionic model \cite{MS} with no memory 
and a three-current ionic model \cite{TSGM} with some memory.
  
\begin{figure}[tbp]
 
  \includegraphics[width=8cm]{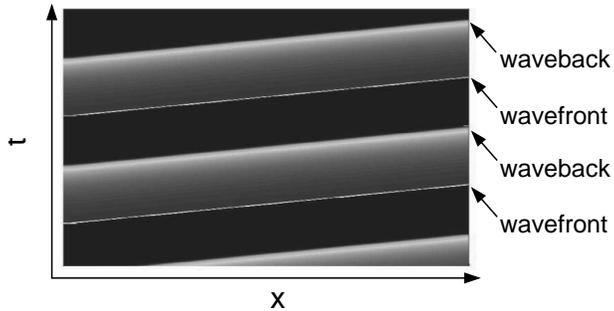}
 
  \caption{Steady-state space-time response obtained by numerical 
  simulation of Eq. \eqref{CABLE} with the two-current ionic model.}
 
  \label{greyscale}

\end{figure}

The details of the numerical experiments are as
follows.  Using a cable of length $L = 10 \: \mathrm{cm}$, we solve 
Eq. \eqref{CABLE} numerically with an operator-splitting method.  Stimuli
are applied over a 
1 mm region at the proximal ($x = 0$) end of the fiber using both the dynamic 
and S1-S2 protocols described in the Introduction.
In all simulations, we use $\Delta = 40$ ms and $\delta = \pm 20$ ms.
Pacing results in a train 
of pulses that propagate left-to-right in the fiber. Measurements of DI, 
APD, wavefront speed, and waveback speed are taken at  
$x = 2.5 \: \mathrm{cm}$.
The position of a pulse wavefront is defined as the $x$ value for which the
transmembrane voltage is $-60 \mathrm{mV}$ and $dv/dx < 0$.
Likewise, waveback position is defined as the $x$ value at which 
$v = -60 \mathrm{mV}$ and $dv/dx > 0$. Linear interpolation is 
used to improve tracking of wavefront and waveback positions. Speeds are 
then computed by recording the time required for wavefronts and wavebacks 
to traverse a 1-mm-wide interval centered at $x = 2.5 \: \mathrm{cm}$.
For illustration purposes, Fig. \ref{greyscale} shows a 
projection of a steady-state solution of Eq. \eqref{CABLE} (with two-current 
ionic model) onto the $xt$ plane.  Different shades of grey correspond to 
different transmembrane voltages, with black corresponding to the rest 
potential.  Note that, in steady-state, projecting the
wavefronts and wavebacks onto the $xt$ plane forms a sequence of parallel
lines.

\begin{figure}[tbp]
 
  \includegraphics[width=8cm]{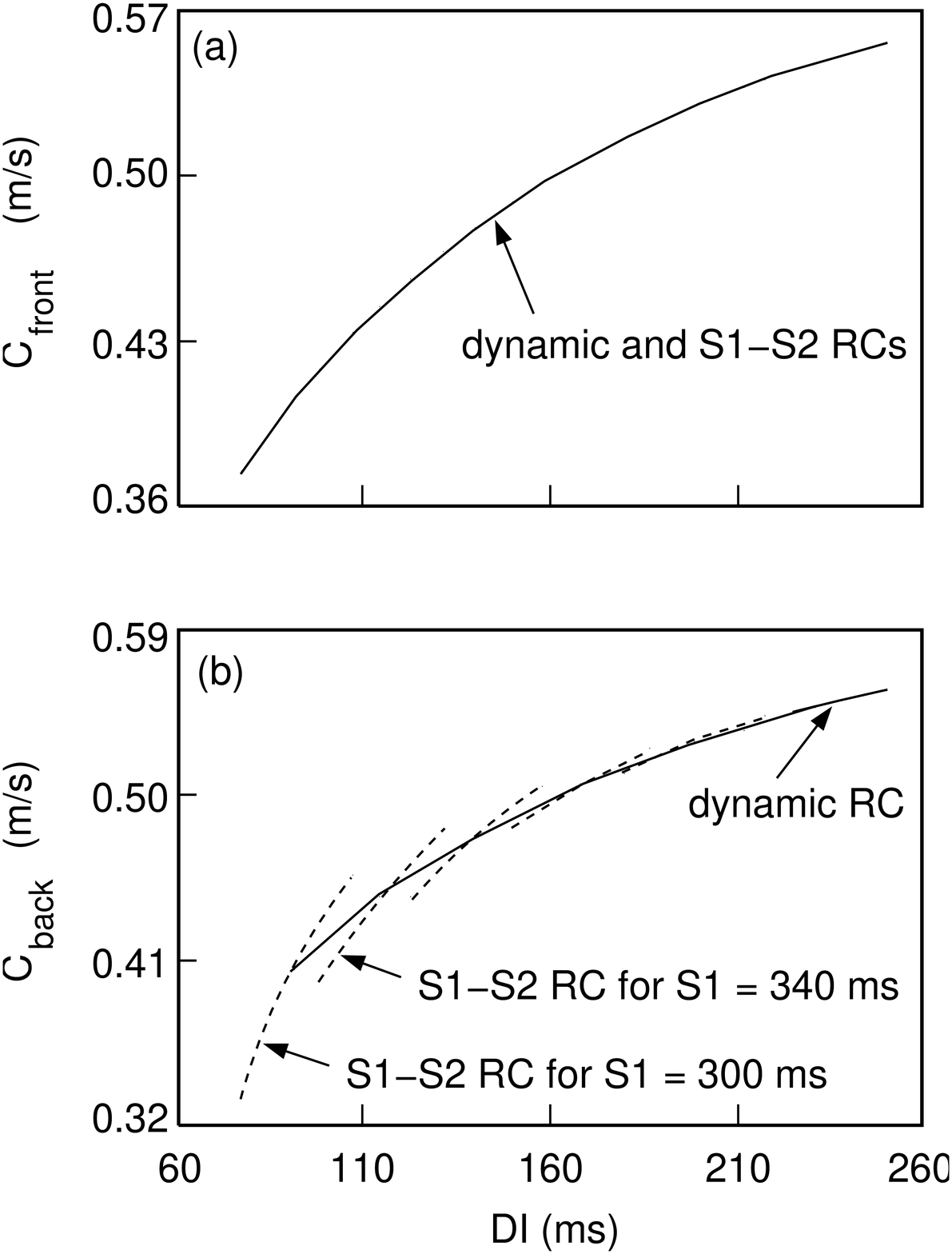}
 
  \caption{Wavefront and waveback velocity RCs obtained 
  from numerical simulation of Eq. \eqref{CABLE} with two-current ionic 
  model ($\Delta = 40 \: \mathrm{ms}$, $\delta = \pm 20 \: \mathrm{ms}$).  
  Velocities were measured at $x = 2.5 \: \mathrm{cm}$. 
  (a) Wavefront velocity RCs.
  (b) Waveback velocity RCs.  The dynamic RC is solid and the local S1-S2
  RCs are dashed.
  }
 
  \label{TCCV}
 
\end{figure}

Results of numerical simulations of Eq. \eqref{CABLE} with the two-current
ionic model (no memory) are presented in Fig. \ref{TCCV}, which shows 
wavefront (Fig. \ref{TCCV}a) and waveback (Fig. \ref{TCCV}b) velocity RCs.
One can see from Fig \ref{TCCV}a that the wavefront velocity
RCs resulting from different pacing protocols are indistinguishable.  Thus,
there is no significant rate-dependence if velocities are measured at the 
wavefront.  However, one can see from Fig \ref{TCCV}b that segments of
S1-S2 waveback velocity RCs (dashed curves) do not coincide with the 
dynamic waveback velocity RC (solid curve).  As in the case of APD
rate-dependence (see Fig. \ref{COMPARE-APD}), the splitting between 
waveback velocity RCs is more pronounced for small values of DI.

\begin{figure}[tbp]
 
  \includegraphics[width=8cm]{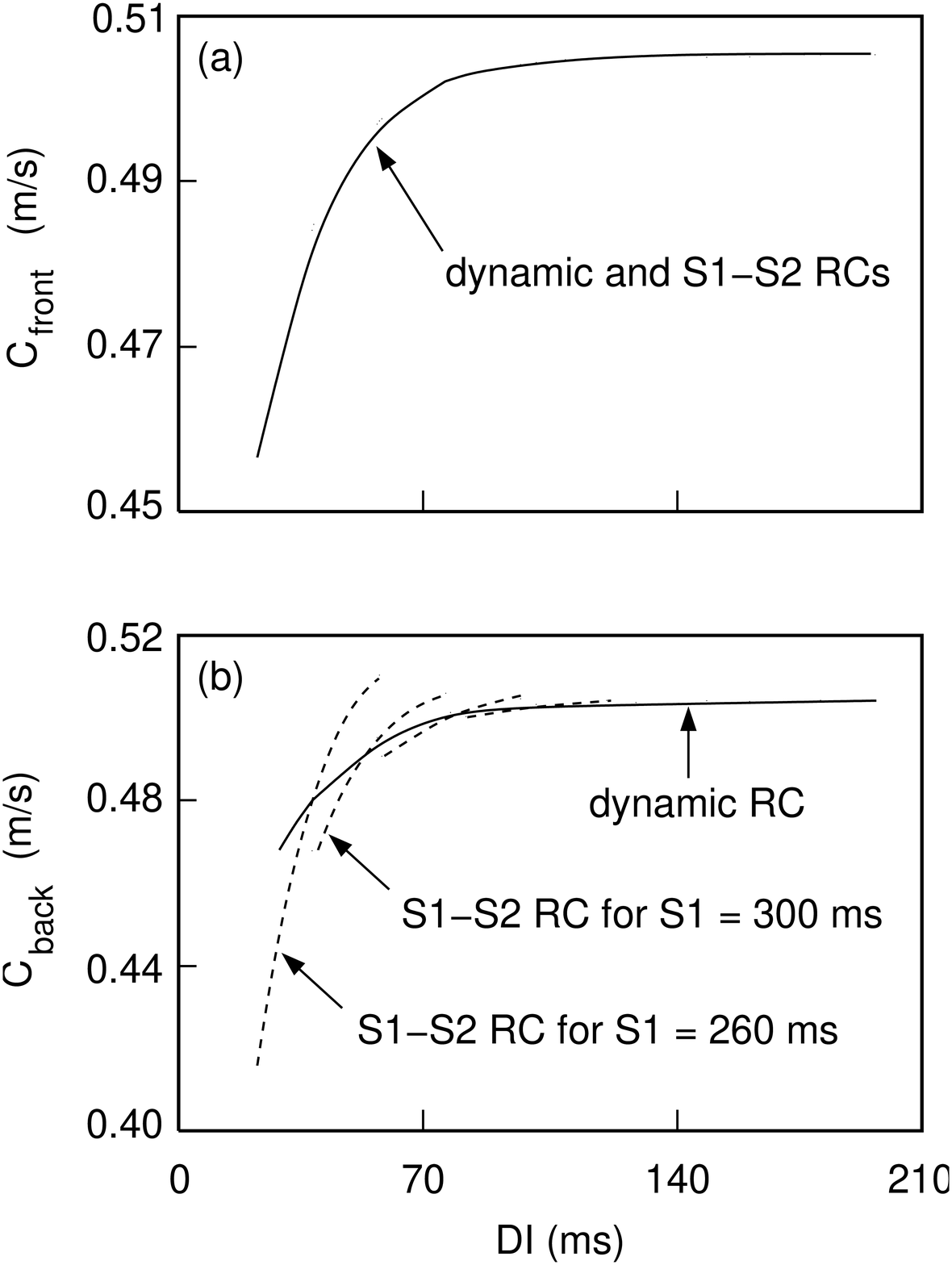}
 
  \caption{Wavefront and waveback velocity RCs obtained 
  from numerical simulation of Eq. \eqref{CABLE} with three-current ionic 
  model  ($\Delta = 40 \: \mathrm{ms}$, $\delta = \pm 20 \: \mathrm{ms}$).  
  Velocities were measured at $x = 2.5 \: \mathrm{cm}$. 
  (a) Wavefront velocity RCs.
  (b) Waveback velocity RCs.  The dynamic RC is solid and the local S1-S2
  RCs are dashed.
  }
 
  \label{FKCV}
 
\end{figure}

Wavefront and waveback velocity RCs obtained from numerical simulations
of a cable with the three-current ionic model (that has some memory) are 
presented in Fig. \ref{FKCV}.  The results are qualitatively similar to the
two-current model results shown in Fig. \ref{TCCV}.

There are two important points that we wish to emphasize.  First, 
rate-dependent waveback velocity restitution does not depend upon the 
presence of memory in the tissue, as evidenced by our two-current model
simulations.  Second, rate-dependent waveback velocity is more pronounced 
for small values of $DI$.  In what follows, we provide an analytical 
explanation of these findings.
 
\section{Rate-dependent velocity: analytical results}
 
Instead of considering the
systems of ODEs that constitute ionic membrane models, many authors
employ mapping models that describe APD as a function of the previous DI and
APD values \cite{ND, TSGM, CHIALVO, FOX}.  To our knowledge, Nolasco and 
Dahlen \cite{ND} were the first to propose a simple mapping model of the 
form
\begin{equation}
  A_{n+1} = F(D_{n})
  \label{NOMEMORY}
\end{equation}
to describe cardiac dynamics.  Here, $A_{n}$ and $D_{n}$ denote the 
$n^{th}$ APD and DI values, respectively.  In general, the number of 
arguments of $F$ determines how much memory is present.

For certain ionic models, such as the two and 
three-current models, it is possible to derive mappings by analyzing 
the ODEs.  As demonstrated in \cite{MS}, a mapping of the form 
\eqref{NOMEMORY} can be derived directly from the two-current model equations
(see Appendix).  The mapping model \eqref{NOMEMORY} has no memory, and 
all APD RCs coincide as in Fig. \ref{COMPARE-APD}a.  

It was shown in \cite{TSGM} 
that the three-current model leads to a mapping with two arguments: 
\begin{equation}
  A_{n+1} = F(A_{n}, D_{n}).
  \label{SOMEMEMORY}
\end{equation}
This mapping model has some memory, and the dynamic and S1-S2 RCs are 
different as in Fig. \ref{COMPARE-APD}b.  
  
In order to explain differences between wavefront and waveback velocity RCs
for different pacing protocols, we approximate the dynamics of Eq.
\eqref{CABLE} with a system of coupled maps.  We follow the approach
described in \cite{WATANABE}, which allows us to derive a relationship between 
wavefront and waveback velocities.  We analyze the dynamic and S1-S2 pacing
protocols separately since the pacing protocol determines the boundary
conditions for Eq. \eqref{CABLE}.

\subsection{Dynamic pacing protocol}

\begin{figure}[tbp]
 
  \includegraphics[width=8cm]{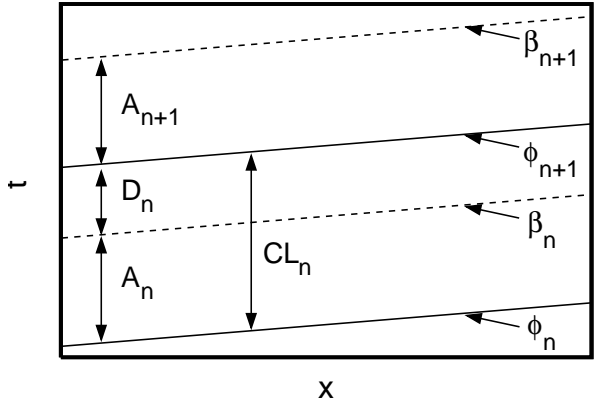}
 
  \caption{Schematic representation of a 1:1 steady-state. Here, 
  $\phi_{n}$ denotes the $n^{th}$ activation front and $\beta_{n}$ is the
  corresponding waveback.  Solid lines represent wavefronts and dashed
  lines represent wavebacks.}
 
  \label{SCHEMATIC}
 
 \end{figure}

Under the dynamic pacing protocol, pacing is performed at a
constant basic cycle length, $B$, at $x = 0$ until steady-state is reached.
In what follows, we assume that a 1:1 steady-state response results from
dynamic pacing.  That is, every stimulus produces an action potential and
there is no beat-to-beat variation in APD or DI.
A schematic representation of steady-state behavior is shown in Fig. 
\ref{SCHEMATIC}, which shows the projection of a particular level set of
the surface in Fig. \ref{greyscale} onto the $xt$ plane.  The lines in 
Fig. \ref{SCHEMATIC} are identified with the sequence of wavefronts
and wavebacks.  We define $\phi _{n}(x)$ ($\beta_{n}(x)$) as the time at 
which the $n^{th}$ wavefront (waveback) reaches $x$.  Note that 
$\phi _{n}(x)$ and $\beta _{n}(x)$ are parallel lines in the $xt$ plane 
if a 1:1 steady-state is reached. The cycle length is defined as
\begin{equation}
  CL_{n}(x)=A_{n}(x)+D_{n}(x)=\phi _{n+1}(x)-\phi _{n}(x).
  \label{DEFINE-CLx}
\end{equation}
 
\noindent 
We remark that $CL_{n}(0)=B$ for all $n$, and $CL_{n}(x) = B$ 
for all $x$ if steady-state is reached.

Let us assume that, at each $x$ along the fiber, APD can be represented as
a function of an arbitrary number of preceding APDs and DIs in a form
\begin{equation}
  \label{MEMORY-VECTORS}
  A_{n+1}(x)=F(\mathbf{A}_{n}(x),\mathbf{D}_{n}(x)),
\end{equation}
i.e. an arbitrary amount of memory is included.  Here,
\begin{eqnarray}
  \label{VECTORS}
  \mathbf{A}_{n}(x) &=&(A_{n}(x),A_{n-1}(x),...,A_{n+1-m}(x)), \\
  \mathbf{D}_{n}(x) &=&(D_{n}(x),D_{n-1}(x),...,D_{n+1-k}(x)), \notag
\end{eqnarray}
and $m \geq 0$ and $k \geq 1$ are integers
characterizing how many preceding states are taken into account in the
mapping model.  Since many previous states are involved, Eq.
\eqref{MEMORY-VECTORS} makes sense only for $m, k \leq n$.
Note that $m=0$, $k=1$ corresponds to the simplest mapping 
model Eq. \eqref{NOMEMORY}, the case of no memory.  The case $m = 1$, $k = 1$
corresponds to a mapping of the form of Eq. \eqref{SOMEMEMORY} with some
memory.

Let us also assume that the velocity of the $(n+1)^{\mathrm{st}}$ wavefront,
$c_{\mathrm{front}}(\mathbf{A}_{n}(x),\mathbf{D}_{n}(x))$, depends upon 
preceding (local) APD and DI values.  This velocity is computed by inverting 
the slope of $\phi_{n}(x)$:
\begin{equation}
  \frac{d \phi_{n+1}(x)}{dx} = 
  \frac{1}{c_{\mathrm{front}}(\mathbf{A}_{n}(x),\mathbf{D}_{n}(x))} \equiv
  G(\mathbf{A}_{n}(x),\mathbf{D}_{n}(x)).
  \label{speed}
\end{equation}
When steady-state is reached, the vectors $\mathbf{A}_{n}(x)$ and
$\mathbf{D}_{n}(x)$ are constant:
\begin{eqnarray}
  \label{DEF-STAR}
  \mathbf{A}_{n}(x) & = & \mathbf{A^{*}} \equiv (\Astar, \Astar, \dots 
  \Astar),\\
  \notag
  \mathbf{D}_{n}(x) & = & \mathbf{D^{*}} \equiv (\Dstar, \Dstar, \dots \Dstar).
\end{eqnarray}
Thus, plotting  
$c_{\mathrm{front}}(\mathbf{A^{*}}, \mathbf{D^{*}})$ versus $\Dstar$, we
obtain a point on the dynamic wavefront velocity RC.  The curves 
$\beta_{n+1}(x)$ and $\phi_{n+1}(x)$ have the same slope
since they are parallel at steady-state: 
$\beta_{n+1}(x) = \phi_{n+1}(x) + \Astar$.
Therefore, the dynamic waveback and wavefront velocity RCs are identical.  
Hence, from now on we refer to {\em the} dynamic velocity RC and use the 
notation $c_{\mathrm{dyn}} = c_{\mathrm{dyn}}(\Dstar)$.
Since propagation speeds typically increase when more recovery is allowed,
we will assume that $c_{\mathrm{dyn}}$ is a monotone increasing function
of $\Dstar$.

It follows from Eqs. \eqref{DEFINE-CLx} and \eqref{speed} that
\begin{equation}
  \frac{d}{dx}CL_{n}(x) = G(\mathbf{A}_{n}(x),\mathbf{D}_{n}(x)) -
  G(\mathbf{A}_{n-1}(x),\mathbf{D}_{n-1}(x)).
  \label{CL-ODE}
\end{equation}
According to Eqs. \eqref{DEFINE-CLx} and 
\eqref{MEMORY-VECTORS}, the cycle length also satisfies an algebraic 
condition
\begin{equation}
  CL_{n}(x) = F(\mathbf{A}_{n-1}(x),\mathbf{D}_{n-1}(x)) + D_{n}(x),
  \label{CL-ALG}
\end{equation}
and thus Eqs. \eqref{CL-ODE} and \eqref{CL-ALG} imply that
\begin{equation}
  \frac{d}{dx} \left[ F(\mathbf{A}_{n-1}(x),\mathbf{D}_{n-1}(x))
  + D_{n}(x) \right] = G(\mathbf{A}_{n}(x),\mathbf{D}_{n}(x))
  - G(\mathbf{A}_{n-1}(x), \mathbf{D}_{n-1}(x)).
  \label{ODESEQ}
\end{equation}
The dynamic pacing protocol gives the following boundary condition at $x=0$:
\begin{equation}
  D_{n}(0) = B -  F(\mathbf{A}_{n-1}(0),\mathbf{D}_{n-1}(0)).
  \label{BC1}
\end{equation}
The sequence of equations Eq. \eqref{ODESEQ} can be solved
iteratively to construct Fig. \ref{SCHEMATIC}.  If the vectors of 
functions $\mathbf{A}_{n}(x)$ and $\mathbf{D}_{n-1}(x)$ are known,  
we can solve \eqref{ODESEQ} to determine $D_{n}(x)$.  Note that 
$A_{n+1}(x)$ can then be computed by applying Eq. \eqref{MEMORY-VECTORS}.

\subsection{S1-S2 pacing protocol}

\begin{figure}[tbp]
 
  \includegraphics[width=8cm]{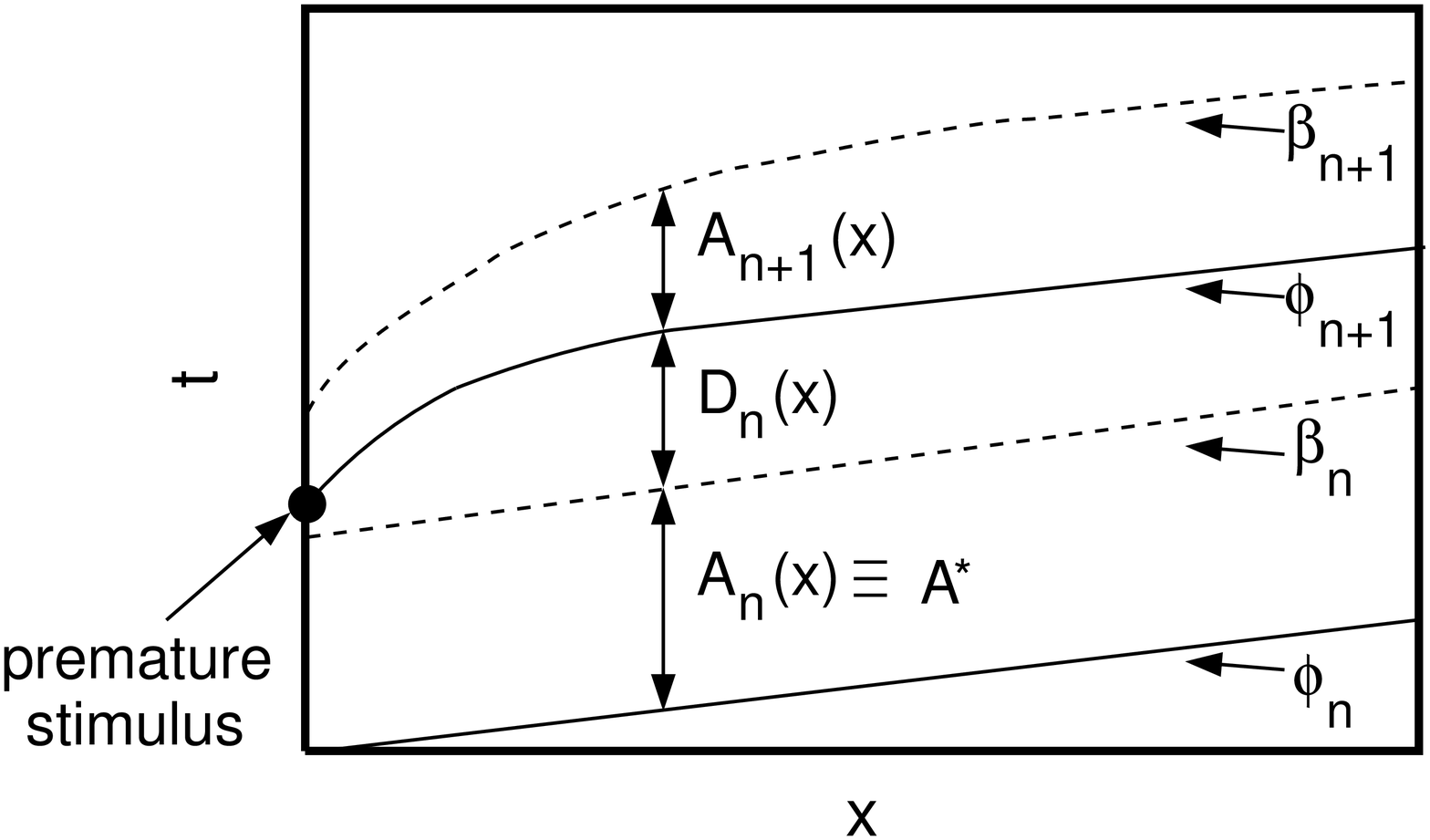}
 
  \caption{Deflection of the $(n+1)^{\mathrm{st}}$ wavefront and waveback 
  due to a premature ($B_{1} < B$) S2 stimulus.  The shortened diastolic 
  interval at the stimulus site slows the propagation speed.  Solid curves
  represent wavefronts and dashed curves represent wavebacks.}
 
  \label{FRONTSANDBACKS}
 
\end{figure}

In the S1-S2 protocol, tissue is paced at a basic cycle length $B$ until
steady-state is reached.  Then, an $S2$ stimulus is introduced at an
interval $B_{1} = B \pm \delta$ following the last S1 stimulus and the
response to the S2 stimulus is measured.  In what follows, we assume that 
the S2 stimulus is applied prematurely ($B_{1} = B - \delta$) following a 
train of $n$ S1 stimuli.  The S2 stimulus causes a deflection in the 
$(n+1)^{\mathrm{st}}$ wavefront and waveback as shown in 
Fig. \ref{FRONTSANDBACKS}.  

These assumptions imply that
\begin{eqnarray}
  \mathbf{A}_{n}(x) &=& \mathbf{A}^{*}, \label{steady}\\
  \mathbf{D}_{n}(x) &=&  (D_{n}(x), \Dstar,..., \Dstar), \notag
\end{eqnarray}
and the boundary condition
\begin{equation}
  D_{n}(0) = B_{1} - \Astar.
  \label{BC2}
\end{equation}
Equation \eqref{ODESEQ} reduces to
\begin{equation}
  \frac{dD_{n}(x)}{dx}=G(\mathbf{A}^{*},\mathbf{D}_{n}(x)) - \Gstar,
  \label{NONLINEAR-IVP1}
\end{equation}
where $\Gstar = G(\mathbf{A}^{*}, \mathbf{D}^{*}) = 
c_{\mathrm{dyn}}^{-1}$.  Linearizing Eq. \eqref{NONLINEAR-IVP1} 
about the point $(\mathbf{A}^{*}, \mathbf{D}^{*})$, we have
\begin{equation}
  \frac{dD_{n}(x)}{dx}= - \lambda \left( D_{n}(x) - D^{*} \right),
  \label{LINEARIZED-IVP}
\end{equation}
where
\begin{equation}
  \lambda = \left. - \frac{\partial {G}}{\partial D_{n}} 
  \right \vert_{(\mathbf{A}^{*},\mathbf{D}^{*})} .
  \label{DEF-LAMBDA}
\end{equation}
Since we assumed that the dynamic velocity RC is monotone increasing, it 
follows that $\lambda > 0$.  The solution of the Eq. \eqref{LINEARIZED-IVP} 
with the boundary condition \eqref{BC2} is
\begin{equation}
  D_{n}(x)= \Dstar - \delta e^{-\lambda x}.
  \label{DN}
\end{equation}

Let $c_{\text{front}}^{S_{12}}$ and $c_{\text{back}}^{S_{12}}$ denote the
wavefront and waveback velocities of the action potential generated by the 
S2 stimulus.  In order to compute $c_{\text{front}}^{S_{12}}$, observe that
(see Fig. \ref{FRONTSANDBACKS})
\begin{equation}
\phi _{n+1}(x) = \beta _{n}(x) + D_{n}(x).
\label{expr0}
\end{equation}
We know that $\phi_{n}(x)$ and $\beta _{n}(x)$ are parallel since they
represent the wavefront and waveback associated with the final S1 stimulus.
Therefore, $d\beta _{n}/dx = d \phi_{n} / dx = \Gstar$, and
differentiating Eq. \eqref{expr0} with respect to $x$ gives
\begin{equation}
  c_{\mathrm{front}}^{S_{12}} = \frac{1}{\Gstar + \delta \lambda  
  e^{-\lambda x}}.
  \label{CFRONT}
\end{equation}
Similarly, to determine $c_{\text{back}}^{S_{12}}$, we use the 
expression (see Fig. \ref{FRONTSANDBACKS}) 
\begin{equation}
  \beta _{n+1}(x)= \phi _{n+1}(x) + A_{n+1}(x) = \phi _{n+1}(x)
  + F(\mathbf{A}_{n}(x), \mathbf{D}_{n}(x)).
  \label{expr2}
\end{equation}
According to Eq. \eqref{steady}, the only non-constant argument of the
function $F$ is $D_{n}(x)$.  Therefore, differentiating Eq. \eqref{expr2} with 
respect to $x$ gives
\begin{equation}
  \frac{d \beta_{n+1}}{dx} = \frac{d \phi_{n+1}}{dx} + 
  \frac{\partial F}{\partial D_{n}}\frac{d D_{n}}{dx} = \Gstar 
  + \delta \lambda  e^{-\lambda x}
  \left( 1 + \frac{\partial F}{\partial D_{n}} \right), 
  \label{differ}
\end{equation}
which implies that
\begin{equation}
  c_{\mathrm{back}}^{S_{12}} = \frac{1}{\Gstar + \delta \lambda  e^{-\lambda x}
  \left( 1 + \frac{\partial F}{\partial D_{n}} \right)}.
  \label{CBACK}
\end{equation}
The partial derivative $\partial F/\partial D_{n}$ in Eq. \eqref{CBACK} is
evaluated at 
$(\mathbf{A^{*}}, D^{*} - \delta e^{-\lambda x}, D^{*}, \dots D^{*})$.
Equations \eqref{CFRONT} and \eqref{CBACK} are analytical expressions for
wavefront and waveback velocity for the S1-S2 pacing protocol.

Both formulas \eqref{CFRONT} and \eqref{CBACK} require that we know 
formulas for the dynamic velocity RC 
(since $\Gstar = c_{\mathrm{dyn}}^{-1}$) and the function $F$.
The only difference between the two formulas is the 
presence of the multiplier 
$\left( 1+\frac{\partial F}{\partial D_{n}}\right)$.  
In simple mapping models for which Eq. \eqref{NOMEMORY} applies, the 
partial derivative $\partial F / \partial D_{n}$ in Eq. \eqref{CBACK} is
replaced by a total derivative $F'(D_{n})$.
Note that the wavefront and waveback velocities approach 
$c_{\mathrm{dyn}}$ as $x\rightarrow \infty$ because $\lambda > 0$.  
If the S2 stimulus is premature ($B_{1} < B$), formulas 
\eqref{CFRONT} and \eqref{CBACK} show that $c_{\mathrm{back}}^{S_{12}} 
< c_{\mathrm{front}}^{S_{12}}$ and the pulse broadens as it propagates.  
Likewise, if the S2 stimulus is late ($B_{1} > B$), then 
$c_{\mathrm{back}}^{S_{12}} > c_{\mathrm{front}}^{S_{12}}$ and the pulse 
contracts as it propagates.  If $B_{1} = B$, the formulas reduce to 
$c_{\mathrm{back}}^{S_{12}} = c_{\mathrm{front}}^{S_{12}}
= c_{\mathrm{dyn}}$ as one would expect.

Equations \eqref{CFRONT} and \eqref{CBACK} reinforce our main point:  
APD restitution, not memory, is responsible for velocity
rate-dependence.  Regardless of how much memory is included in the
mapping model, Eqs. \eqref{CFRONT} and \eqref{CBACK} depend upon $D_{n}(x)$ 
and no other preceding states.  
The partial derivative $\partial F / \partial D_{n}$ in Eq. \eqref{CBACK}
represents the slope $S_{12}$ of the S1-S2 APD RC as demonstrated in 
\cite{TSGK}.
As $DI$ decreases, $S_{12}$ typically increases, thereby increasing the 
discrepancy between the wavefront and waveback velocities.  It follows that, 
in the absence of wavefront velocity rate-dependence, APD restitution 
leads to waveback velocity rate-dependence.

\subsection{An example: Rate-dependent velocity and the two-current model}

In this Subsection, we explain how to apply Eqs. \eqref{CFRONT}
and \eqref{CBACK}, using the two-current model as an example.  
As mentioned above, Eqs. \eqref{CFRONT} and \eqref{CBACK} require that we 
provide formulas for the function $F$ and the dynamic velocity RC.  
Leading-order expressions for $F$ and $c_{\mathrm{dyn}}$ can be 
derived analytically for the two-current model (see Appendix).

The dynamic velocity RC is provided by Eq. \eqref{LEADING-C}.
Combining Eqs. \eqref{CFRONT} and \eqref{LEADING-C}, we generate all S1-S2
wavefront velocity RCs.  Likewise, combining Eqs. \eqref{CBACK},
\eqref{LEADING-F}, and \eqref{LEADING-C} allows us to construct all S1-S2
waveback velocity RCs.

\begin{figure}[tbp]
 
  \includegraphics[width=8cm]{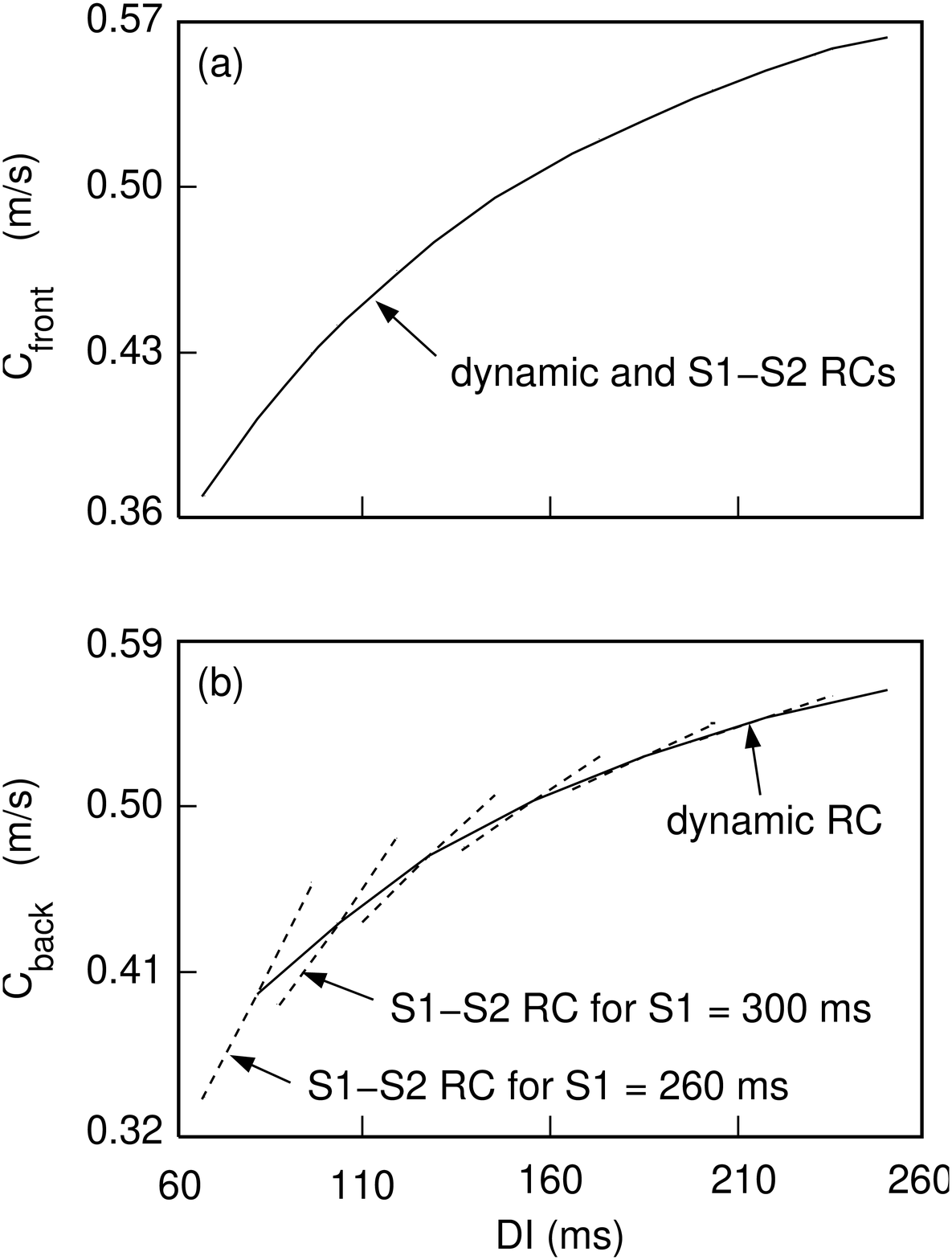}
 
  \caption{Wavefront and waveback velocity RCs generated using Eqs.
  \eqref{CFRONT}, \eqref{CBACK}, \eqref{LEADING-F}, and \eqref{LEADING-C}.
  ($\Delta = 40 \: \mathrm{ms}$, $\delta = \pm 20 \: \mathrm{ms}$). 
  Velocities were measured at $x = 2.5 \: \mathrm{cm}$. 
  (a) Wavefront velocity RCs: the dynamic and S1-S2 curves appear to coincide.
  (b) Waveback velocity RCs: the dynamic curve is solid and local 
  S1-S2 curves are dashed.
  }
 
  \label{ANALYTICAL-TC}
 
\end{figure}

All of the analytically derived RCs are shown in Fig. \ref{ANALYTICAL-TC}.  
Figure \ref{ANALYTICAL-TC}a shows all wavefront velocity RCs.  The
dynamic and S1-S2 wavefront velocity RCs are indistinguishable.  The 
waveback velocity
RCs are shown in Figure \ref{ANALYTICAL-TC}b.  Note the presence of
rate-dependence, as evidenced by the splitting of the dynamic (solid) and
S1-S2 (dashed) RCs.  We remark that Fig. \ref{ANALYTICAL-TC} 
shows excellent quantitative agreement with the results of numerical 
simulations shown in Fig. \ref{TCCV}.

\section{CONCLUSIONS}

We have demonstrated that rate-dependent waveback velocity restitution can
exist even in the absence of memory. Our numerical simulations show that
both the two and three-current models exhibit rate-dependent waveback 
velocity, whereas neither model exhibits rate-dependent wavefront
velocity.  We offer a mathematical explanation for the
differences between wavefront and waveback dynamics.  Specifically, 
comparison of Eqs. \eqref{CFRONT} and
\eqref{CBACK} shows that as the slope $S_{12}$ of the S1-S2 APD RC increases,
the difference between the wavefront and waveback velocities is
magnified.  Therefore, if the S1-S2 wavefront velocity RCs coincide with
the dynamic velocity RC, then the S1-S2 waveback velocity RCs cannot.
Moreover, Eqs. \eqref{CFRONT} and \eqref{CBACK} predict that
the splitting between the dynamic and S1-S2 waveback velocity
RCs should be most pronounced at small $DI$ values where $S_{12}$ is
largest.  These analytical predictions are consistent with the results of 
our numerical experiments.  Finally, the validity of the computations in 
Sec. III is strongly supported by the quantitative agreement between 
numerical and analytical investigations of the two-current model
(Figs. \ref{TCCV} and \ref{ANALYTICAL-TC}).


\begin{acknowledgments}

We gratefully acknowledge the financial support of the National
Science Foundation through grants PHY-0243584 and DMS-9983320 and the
National Institutes of Health through grant 1R01-HL-72831.

\end{acknowledgments}

 
\appendix*

\section{}

A detailed analysis of the two-current model equations appears in \cite{MS}. 
With the two-current model,  Eq. \eqref{CABLE} reads
\begin{eqnarray} 
  \label{CABLE-TC}
  \frac{\partial v}{\partial t} & = & \kappa \frac{\partial^{2} v}{\partial
  x^{2}} + \frac{h}{\tau_{\mathrm{in}}} v^{2} \left( 1 - v \right) 
  - \frac{v}{\tau_{\mathrm{out}}} \\
  \label{DHDT}
  \frac{dh}{dt} & = & 
  \begin{cases}
  \frac{1 - h}{\tau_{\mathrm{open}}} & v < v_{\mathrm{crit}} \\ 
  - \frac{h}{\tau_{\mathrm{close}}} & v > v_{\mathrm{crit}},
  \end{cases}
\end{eqnarray}
where $v$ is transmembrane voltage (scaled to range between 0 and 1) and $h$
is a gate variable.  The parameters $\tau_{\mathrm{in}}$, 
$\tau_{\mathrm{close}}$, $\tau_{\mathrm{out}}$, and $\tau_{\mathrm{open}}$ 
are time constants associated with different phases of the action potential. 
The gate opens or closes according to whether $v$ exceeds the threshold 
voltage $v_{\mathrm{crit}}$.  Typical choices for the time constants and
critical voltage are: $\tauin = 0.1$ ms, $\tauout = 2.4$ ms,
$\tauopen = 130$ ms, $\tauclose = 150$ ms, and $\vcrit = 0.13$.

A leading-order estimate of the APD RC is derived in \cite{MS}.  
If the time constants satisfy an asymptotic condition 
$\tau_{\mathrm{in}} \ll \tau_{\mathrm{out}} \ll \tau_{\mathrm{open}}, 
\: \tau_{\mathrm{close}}$, then
\begin{equation} 
  A_{n+1} = F(D_{n}) = \tau_{\mathrm{close}} \ln 
  \left( \frac{h_{s}(D_{n})}{h_{\mathrm{min}}} \right)
  \label{LEADING-F}
\end{equation}
to leading order, where
\begin{equation} 
  h_{s}(D_{n}) = 1 - \left(1 - h_{\mathrm{min}} \right) 
  e^{-\frac{D_{n}}{\tau_{\mathrm{open}}}},
  \label{HSTAR}
\end{equation}
and $h_{\mathrm{min}} = 4 \tau_{\mathrm{in}} / \tau_{\mathrm{out}}$.
 
To derive a leading-order estimate of $c_{\mathrm{dyn}}$, we follow 
Murray \cite{MURRAY}.  Assume the fiber is paced at a constant basic 
cycle length until steady-state is reached so that all pulses
propagate with speed $c_{\mathrm{dyn}} = c_{\mathrm{dyn}}(\Dstar)$.
We seek traveling wavetrain solutions to Eq. \eqref{CABLE-TC}.  In the
neighborhood of a wavefront, introduce the coordinate
\begin{equation}
  \xi = \frac{1}{\tauin} \left( t + \frac{x}{c_{\mathrm{dyn}}}
  \right),
  \label{DEF-XI}
\end{equation}
where the speed $c_{\mathrm{dyn}}$ is to be determined. Assume
that $v(x,t) = V(\xi)$ and $h(x, t) = H(\xi)$. Since $\tau_{\mathrm{in}}$ is
small relative to the time constants in Eq. \eqref{DHDT}, we
may safely approximate the value of $h$ by a constant in the narrow
wavefront region:  $h \approx h^{*} \equiv h_{s}(\Dstar)$.  Inserting 
$v(x,t) = V(\xi)$
into Eq. \eqref{CABLE-TC}, we obtain an ordinary differential equation
\begin{equation} 
  \frac{\kappa}{c_{\mathrm{dyn}}^{2} \tau_{\mathrm{in}}} 
  V'' - V' + h^{*} \: V \left( V_{-} - V \right) 
  \left(V - V_{+} \right) = 0,
  \label{TRAVELING}
\end{equation}
where primes denote differentiation with respect to $\xi$ and
\begin{equation} 
  V_{\pm} = \frac{1}{2} \left( 1 \pm \sqrt{ 1 - 
  \frac{h_{\mathrm{min}}}{h^{*}}} \right).
  \label{VPM}
\end{equation}
We remark that $V_{-}$ is an unstable equilibrium of 
Eq. \eqref{TRAVELING} corresponding to the threshold for excitation, and
$V_{+}$ is an unstable equilibrium associated with the excited state.
We seek solutions to Eq. \eqref{TRAVELING} such that $V(\xi) \rightarrow 0$ as 
$\xi \rightarrow -\infty$ and $V(\xi) \rightarrow V_{+}$ as 
$\xi \rightarrow \infty$. It is possible to find a solution of a simpler 
differential equation
\begin{equation}
  V^{\prime}= -a V \left( V - V_{+} \right)
  \label{SIMPLER-ODE}
\end{equation}
that also satisfies Eq. \eqref{TRAVELING} for unique values of the constant $a$
and the speed $c_{\mathrm{dyn}}$.  Substituting \eqref{SIMPLER-ODE}
into Eq. \eqref{TRAVELING}, one finds that
\begin{equation} 
  c_{\mathrm{dyn}} = \left( \frac{1}{2} V_{+} - V_{-}  \right) 
  \sqrt{\frac{2 \kappa h^{*}}{\tau_{\mathrm{in}}}}.
  \label{LEADING-C}
\end{equation}


 \end{document}